\documentclass[man,apacite,floatsintext]{apa6}
\usepackage{epsfig,amsmath,alltt,bm,url,subcaption, changes}
\usepackage[english]{babel}

\title{Self-normalized score-based tests to detect parameter heterogeneity for mixed models}
\twoauthors{Ting Wang}{Edgar C.\ Merkle}
\twoaffiliations{American Board of Family Medicine}{University of Missouri}

\abstract{
Score-based tests have been used to study parameter heterogeneity across many types of statistical models. This chapter describes a new self-normalization approach for 
score-based tests of mixed models, which addresses situations where there is dependence between scores. This differs from the traditional score-based tests, which require independence of scores. We first review traditional score-based tests and then propose a new, self-normalized statistic that is related to previous work by \citeA{shao10} and \citeA{zhang11}. We then provide simulation studies that demonstrate how traditional score-based tests can fail when scores are dependent, and that also demonstrate the good performance of the self-normalized tests. Next, we illustrate how the statistics can be used with real data. Finally, we discuss the potential broad application of self-normalized, 
score-based tests in mixed models and other models with dependent observations.}

\authornote{This paper was partially supported by NSF grant 1460719. Thanks to Professor Xianyang Zhang for sharing the original paper's code, and thanks for Rudolf Debelak for comments that helped improve the paper. All remaining errors are the authors' responsiblity.  
  Correspondence to Ting Wang. Email: szpku.grady@gmail.com}
\shorttitle{Self-normalized scores}
\rightheader{Self-normalized scores}

\newcommand{\argmax}{\operatorname{argmax}\displaylimits}
\let\proglang=\textsf
\let\pkg=\emph

\setkeys{Gin}{width=\textwidth}

\usepackage{Sweave}
\begin{document}

\include{Sweave}
\Sconcordance{concordance:chapter.tex:chapter.Rnw:%
1 33 1 1 0 7 1 1 17 270 1 1 33 7 1 1 19 1 2 7 1 1 18 1 2 1 1 1 37 7 1 1 %
18 1 2 8 1 1 18 1 2 1 1 1 5 9 1 1 15 17 0 1 2 3 1 1 5 9 1 1 15 17 0 1 2 %
24 1 1 13 3 1 1 2 1 0 2 1 1 2 4 0 1 2 18 1 1 3 5 0 1 2 16 1 1 22 3 1 1 %
2 1 0 3 1 2 2 3 1 3 0 1 2 4 1 1 9 1 2 62 1}

\maketitle



Score-based tests have been used to study parameter heterogeneity in 
a variety of fields including economics \cite<e.g.,>{and93, hjo02}, policy analysis \cite<e.g.,>{zei07}, and psychometrics \cite<e.g.,>{debstr19, merzei13, merfanzei, schstr22, wang18}.  
The essential idea is to detect 
whether one or more model parameters vary with an unmodeled, auxiliary variable. This approach has the advantage of  simplifying the statistical model of interest (auxiliary variables are not included in the model) and providing information about values of the auxiliary variable for which model parameters differ.
For example, in the context of factor analysis, we can use score-based tests to study measurement non-invariance. The tests help us study whether or not factor loadings vary with respect to a continuous variable, such as respondent age. If the loadings do vary, then the tests also provide information about age groups for which loadings differ \cite{merzei13}. 
Additionally, if there exist multiple auxiliary variables (more than just ``age''),
researchers can consider tree approaches that use score-based tests to select important auxiliary variables \cite<e.g.,>{arnvoe21,bravon13,fok15,hendeb22,strobl15,zei08}.

The above applications almost always involve the assumption that the scores entering in to the test are independent. Such an assumption is inherited for regression and generalized linear models, where cases are assumed to be independent. But the assumption can be violated in mixed models, where cases within the same group (cluster) are correlated. If we ignore this dependence, then the score-based tests will exhibit suboptimal statistical properties \cite<low power, incorrect Type I error rates;>{wang18}. 

A small number of methods have been proposed to handle score dependence in mixed models. For models with nested random effects, we can aggregate the scores to the highest level so that they are again independent \cite<e.g.,>{wang21}. This aggregation to the highest level implicitly happens in many psychometric applications, where the highest level is the person. But it is restrictive because we can only study parameter heterogeneity with respect to the highest level of the data. To address this issue, \citeA{fok15} propose a two-step procedure where a generalized linear mixed model is first estimated, then the model is ``downgraded'' to a generalized linear model by conditioning on predictions of the random effects. The scores are independent in this downgraded model, so that the score-based tests can be carried out as usual. While this solution is sufficient for many applications, it also ignores uncertainty in the random effect predictions and cannot be used to study heterogeneity in random effect (co)variance parameters. It seems worthwhile to find a solution that can be applied to all mixed models and that does not require us to condition on random effects.

In this chapter, we apply a self-normalization procedure \cite{shao10b,shao10, zhang11} to handle case dependence in score-based tests for linear mixed models. The self-normalization procedure was previously applied to detect change points in time series analysis, where the self-normalization handles unmodeled dependence between cases. That is, instead of assuming a particular autoregressive or moving average structure, the self-normalization procedure allowed the data to ``speak for themselves'' when accounting for dependence and studying change points. We use the same procedure here, where we do not attempt to use any analytic results pertaining to case dependence in mixed models. Instead, we use self-normalization to characterize the empirical dependence between cases, which allows us to decorrelate the scores.

In the current chapter, we first provide specific background on 
score-based tests and on the self-normalization procedure. We then conduct simulation studies to compare traditional score-based tests with self-normalized, score-based tests. Next, we apply the self-normalized tests to empirical data. Finally, we discuss the potential broad application of the self-normalized, 
score-based tests in linear mixed models and other models where cases exhibit dependence.

\section{Score-based tests}

   
This section includes a brief introduction to score-based tests; for a more detailed introduction, see \citeA{wanmer14}.

In the usual maximum likelihood estimation framework, the goal is to find the 
parameter vector $\hat{\bm \theta}$ (say, of length $q)$ that maximizes the model's log likelihood
$\ell$: 
\begin{equation}
  \label{eq:theta}
  \hat{\bm \theta}=\argmax_{\bm \theta} \ell(\bm \theta; \bm{y}_1, \ldots, \bm{y}_n), 
\end{equation}
where $\bm{y}_1, \ldots, \bm{y}_n$ represents the observed data.  
Maximization of the likelihood function 
is equivalent to finding $\hat{\bm \theta}$ such that the \emph{score} vectors satisfy 
the following condition: 
\begin{equation}
  \label{eq:scoresum}
  \sum_{i=1}^{n}s(\hat{\bm \theta}; \bm y_i) = \bm{0}, 
\end{equation}
where a \emph{score} is the partial first derivative of the model's log 
likelihood for case $i$:  
\begin{equation}
  \label{eq:score}
  s(\bm \theta; \bm y_i) = \left(\frac{\partial \ell(\bm \theta;\bm y_i)}{
  \partial \theta_1}, \ldots, \frac{\partial \ell(\bm \theta; \bm y_i)}{
  \partial \theta_j}  \right)^{T}. 
\end{equation}

Score-based tests are used to study how model parameters fluctuate with unmodeled, auxiliary variables. To carry out a test, we organize the score vectors into a scaled, cumulative score process defined as:
\begin{equation} 
    \label{eq:cumscore}
  {\bm B}(t; \hat {\bm \theta}) ~=~ n^{-1/2} \hat {\bm I}^{-1/2} 
    \sum_{i = 1}^{\lfloor n \cdot t \rfloor} {\bm s}(\hat {\bm \theta}; \bm{y}_{(i)}), 
\end{equation}
where $t$ is in $[0,1]$; $\bm{y}_{(i)}$ represents the $i$th-smallest observation (ordered with respect to the unmodeled auxiliary variable);
and $\hat{\bm I}$ represents the information matrix, which has an analytical 
expression for linear mixed models \cite<see>[Section 3.3]{wang182}. 
The purpose of $\hat{\bm I}$ is to decorrelate the cumulative score matrix's columns. In other words, a case's score with respect to one parameter is typically correlated with its score with respect to other parameters. Premultiplication by $\hat {\bm I}^{-1/2}$ decorrelates these scores, which facilitates development of score-based tests. 

Using this scaled cumulative score process, multiple test statistics can be computed.
We will focus on one, the {\em Cram\'{e}r von Mises} (CvM) statistic (named after two mathematicians), because it is similar to the self-normalized statistic that we describe later. The CvM statistic averages the sum of squared values of the cumulative scores, across all points of the cumulative score process. 
To facilitate comparison to the proposed statistic, the CvM statistic can be written as: 
\begin{eqnarray}
  \label{eq:transfer1}
   \mathit{CvM} & = & n^{-1} \sum_{k = 1}^n {\bm B(k/n; \hat{\bm \theta})}^{T}
          {\bm B (k/n; \hat{\bm \theta})} \\
     \label{eq:transfer2}
                & = & n^{-1} \sum_{k = 1}^n \left [ \left \{n^{-1/2} \hat {\bm I}^{-1/2} 
    \sum_{i = 1}^{k} {\bm s}(\hat {\bm \theta}; y_{(i)})\right \}^{T}
          \left \{n^{-1/2} \hat {\bm I}^{-1/2} 
    \sum_{i = 1}^{k} {\bm s}(\hat {\bm \theta}; y_{(i)})\right \} \right ] \\
      \label{eq:transfer3}
                & = & n^{-2} \sum_{k = 1}^n \left [ \left \{ 
                 \sum_{i = 1}^{k} {\bm s}(\hat {\bm \theta}; y_{(i)}) \right \}^{T} \hat {\bm I}^{-1} 
                 \left \{\sum_{i = 1}^{k} {\bm s}(\hat {\bm \theta}; y_{(i)})\right \} \right ].
\end{eqnarray}
We further define partial cumulative scores as
\begin{align}
  \bm B^{\star}_{a, b} &= \displaystyle \sum_{i = a}^{b} {\bm s}(\hat {\bm \theta}; y_{(i)})\ \ \  a \leq b \\
  \bm B^{\star}_{a, b} &= \displaystyle \sum_{i = b}^{a} {\bm s}(\hat {\bm \theta}; y_{(a + b - i)})\ \ \  a > b.
\end{align}
The first line, where $a \leq b$, is the usual cumulative sum starting at the score associated with $y_{(a)}$ and ending at the score associated with $y_{(b)}$. The second line, where $a > b$, is a backward cumulative sum starting with the score associated with $y_{(a)}$ and ending at the score associated with $y_{(b)}$. Using this notation, we can rewrite Equation~\eqref{eq:transfer3} as
\begin{equation}
  \label{eq:transfer4}
\mathit{CvM} = n^{-2} \sum_{k = 1}^n  \bm{B}^{\star^T}_{1,k} \hat {\bm I}^{-1} \bm B^\star_{1,k}.
\end{equation}
This form of the CvM statistic will be helpful to keep in mind, as we consider the self-normalization statistic below.

\section{Self-Normalization}
While the information matrix decorrelates column-wise dependence among scores (i.e, dependence between parameters), it does not address row-wise dependence (between cases). As mentioned earlier, this is a problem for mixed modeling, where cases within the same cluster are correlated. If we applied the traditional, score-based test statistics to situations where row-wise dependence exists, the resulting test statistic will be incorrect \cite{per06}. \citeA{and91} proposed substitutes for $\hat{\bm I}$ in this situation, but these substitutes introduce an extra tuning parameter, which is difficult to implement in practice.

The method that we study here comes from a series of papers by Shao and Zhang \cite{shao10, zhang11}, who proposed a self-normalization approach for time series data that is related to previous work by \citeA{lob01} and others \cite{kie00,kie05}.  A key idea is to further decorrelate dependence of the scores as a function of $k$, where $k \in {1, \ldots, n-1}$ represents each possible changing point.  
Formally, the statistic is defined as: 
\begin{equation}
  \label{eq:dm2}
  SN = \sup_{k= 1, \ldots, n-1} \bm{T}_n(k) 
             ^{T} \bm V_{n}^{-1}(k) 
             \bm {T}_n(k),
\end{equation}
where $\bm{T}_n(k)$ and $\bm{V}_n(k)$ are defined below.

The $\bm T_{n}(k)$ vector is the difference between the cumulative score process at point $k$ and its expected value. This can be written as
\begin{equation}
  \label{eq:t}
  \bm{T}_n(k) = n^{-1/2}\left (\bm B^{\star}_{1, k} - \frac{k}{n} \bm B^{\star}_{1, n} \right).
\end{equation}
Because the sum of scores across all cases is by definition zero (e.g., $\bm B^{\star}_{1, n} = \bm{0}$), the second term in parentheses disappears in most score-based applications. The only exception here may be situations where the scores involve an integral approximation, as in GLMMs \cite<e.g.,>{wangra22}. In these situations, the scores may not sum to exactly zero due to the approximation.

Next, $\bm V_n(k)$ involves the (co)variance in the cumulative scores, considered separately before and after point $k$. Before point $k$ we compute covariances between the usual cumulative score, whereas after point $k$ we compute covariances between the backward cumulative score that starts at point $n$ and ends at point $k+1$.

As an intermediate step toward defining $\bm V_n(k)$, let $\bm C_{a,b}$ be a matrix containing deviations of the cumulative sum $\bm B^\star$ from its expected value between points $a$ and $b$. Specifically, for $a \leq b$, we have a $q \times (b - a + 1)$ matrix $\bm C_{a,b}$ whose $j$th column is
\begin{equation*}
  \bm B^\star_{a,(a+j-1)} - \frac{j}{(b - a + 1)} \bm B^\star_{a,b}
\end{equation*}
for $j = 1, \ldots, (b - a + 1)$. Conversely, for the backward cumulative sum with $a > b$, $\bm C_{a,b}$ is a $q \times (a - b + 1)$ matrix whose $j$th column is
\begin{equation*}
  \bm B^\star_{a,(a-j+1)} - \frac{j}{(a - b + 1)} \bm B^\star_{a,b} \ \ \ a > b
\end{equation*}
for $j = 1, \ldots, (a - b + 1)$. 
Using these intermediate matrices, we write $\bm{V}_n(k)$ as
\begin{equation}
  \label{eq:v}
  \bm{V}_n(k) = n^{-2}\left [\bm C_{1,k} \bm C_{1,k}^T + \bm C_{n,k+1} \bm C_{n,k+1}^T \right ],
\end{equation}
which includes the covariance of entries of the cumulative scores up to point $k$, along with the covariance of entries of the backward cumulative scores down to point $k+1$. This form of matrix was developed for vector time series applications where the autocovariance function is unknown. Multiplication by $n^{-2}$ facilitates computation of the resulting statistic's null distribution. The $\bm{V}_n(k)$ matrix basically is chosen so that, when we take its inverse, it cancels out the covariance that exists in the cumulative scores $\bm {T}_n(k)$. See \citeA{shao15} for further discussion and alternate expressions for $\bm{V}_n(k)$.
\ \\ \ \\

Comparing the CvM from Equation~\eqref{eq:transfer4} to the self-normalization statistic from Equation~\eqref{eq:dm2}, we can observe three main differences. First, the self-normalization statistic involves the dynamic matrix $\bm{V}_n(k)$ that changes with $k$ and that can decorrelate (``self-normalize'') both row and column dependency in the cumulative score matrix. Second, the summing operation from the CvM statistic has changed into a maximum ($\sup$). It seems possible that we could also use a summing operator in the self-normalization procedure, but we do not consider it here. We return to this issue in the General Discussion. Third, the multiplication by $n$ differs across the two statistics. This is not a crucial difference because critical values of the $SN$ statistic are obtained via simulation, as described next.


The $SN$ statistic converges to a function of a Brownian bridge 
\cite<for details, see>{shao10,zhang11}. We realize that some readers may be unfamiliar with a Brownian bridge, and we refer those readers to \citeA{wanmer14} for further discussion and additional references. For our purposes here, it is sufficient to know that the Brownian bridge is a well-studied statistical process that is used in other score-based tests and that allows us to derive critical values and p-values under the null hypothesis.
For the $SN$ statistic defined above, we can compute critical values by repeatedly simulating the Brownian bridge and computing the $SN$ statistic. This is similar to how some other score-based test statistics are obtained \cite<e.g.,>{merfanzei}.



In the following section, we will demonstrate problems with applying traditional score-based tests to correlated cases, and we will study the performance of the self-normalized test that we just described. 

\section{Simulation}
In two-level, linear mixed models, observations in the same cluster are correlated with one another.  This causes problems for traditional score-based tests but not necessarily for self-normalized tests. We now illustrate these issues via simulation and the simulation code is available at \url{http://semtools.r-forge.r-project.org/}.

\subsection{Method}
The simulation setup is similar to \citeA{wang21}, with the only 
difference being that the level 1 scores are being used in place of aggregated level 2 scores (where the level 2 scores are independent of each other).  Our model is similar to that used for the \emph{sleepstudy} data \cite{bel03} that are included with \pkg{lme4} \cite{lme4}. These data come from a longitudinal study of the association between sleep deprivation and reaction time, with each 
subject contributing $10$ observations. The dependent variable is reaction time, and the predictor is 
\emph{Days} of sleep deprivation. The fixed effects are therefore an intercept, $\beta_0$, and a slope for \emph{Days}, $\beta_1$.   
We also have covarying intercept and slope random effects, leading to two random effect variances ($\sigma_0^2$, $\sigma_1^2$) and one covariance ($\sigma_{01}$).  The variance not captured by 
the random effects is modeled by the residual 
variance $\sigma_r^2$. The model can be expressed as: 
\begin{eqnarray}
  \label{eq:prob1}
  \text{Reaction Time}_j \mid\ b_{0j}, b_{1j} &\sim& N(\beta_{0} + b_{0j} + (\beta_{1} + b_{1j}) Days , \bm R_j)\\
  \label{eq:prob2}
  \left( \begin{array}{c} b_{0j} \\ b_{1j} \end{array} \right ) &\sim& N \left ( \bm 0, \left[ {\begin{array}{cc}
   \sigma_0^2 &  \sigma_{01}\\
   \sigma_{01} & \sigma_{1}^2\\
  \end{array} } \right] \right) \\
  \label{eq:prob3}
  \bm R_j &=& \sigma_r^2\bm I_{10}, 
\end{eqnarray}
where $j$ indicates subject $j$, $Days$ is a vector containing the values 0 to 9, and $\bm I_{10}$ denotes identity matrix with dimension of 10. 

For simplicity, 
we focus on instability in fixed effect parameters with respect to an unmodeled continuous variable, which we loosely call {\em cognitive ability}. In one condition, the fixed intercept differs for individuals below the median level of cognitive ability, as compared to individuals above the median level of cognitive ability. In a second condition, the fixed slope differs instead of the fixed intercept.
The magnitude of differences between parameters is based on a $d$ parameter. When $d$ is 0, it 
represents no change in the corresponding parameter, which serves as the 
baseline; when $d$ is greater than 0, it 
represents differences in parameter values, with larger $d$ indicating more severe 
parameter change. In this simulation, $d \in {0, 1, 2, 3, 4}$, which roughly indicates the number of standard errors by which parameters differ across groups. 

The data generating model's true parameter values were set to be the same as the estimates from 
the \emph{sleepstudy} data.
Model estimation proceeded via marginal maximum likelihood, where the form of the data generating model matched that of the fitted model (except for parameter differences between groups).  Parameter instability was tested individually in the fixed intercept parameter and in the fixed slope parameter. 

For both tests, we examine power and Type I error across three sample 
sizes (24, 48, or 96 individuals, contributing a total of $n = 240, 480,$ or $960$ observations), five magnitudes of parameter change, and two parameter conditions (change in either the fixed intercept or fixed slope).  
For each combination of conditions, we generate
$1,000$ data sets. 
For traditional score-based tests, 
we examine three continuous statistics. Specifically, in addition to the $\text{CvM}$ statistic mentioned above, 
we also compute the double maximum statistic ($\text{DM}$) and maximum Lagrange multiplier test ($\text{maxLM}$). 
These two statistics differ from $\text{CvM}$ on 
different aggregation strategies for the cumulative score matrix $\bm B(t; \bm \theta)$; for more details, see \citeA{merzei13}. 
The main focus of this simulation is to compare all of these statistics to the self-normalized 
statistic proposed in this chapter. 

\subsection{Results}
\begin{figure}[H]
\caption{\scriptsize{Simulated power curves for $\text{CvM}, \text{DM}, \text{maxLM}$ 
          across parameter change of
         0--4 asymptotic standard error. 
         The changing parameter is $\beta_{0}$. Panel labels denote
         the parameter being tested along with sample size.}}
\label{fig:sim11res}
\includegraphics{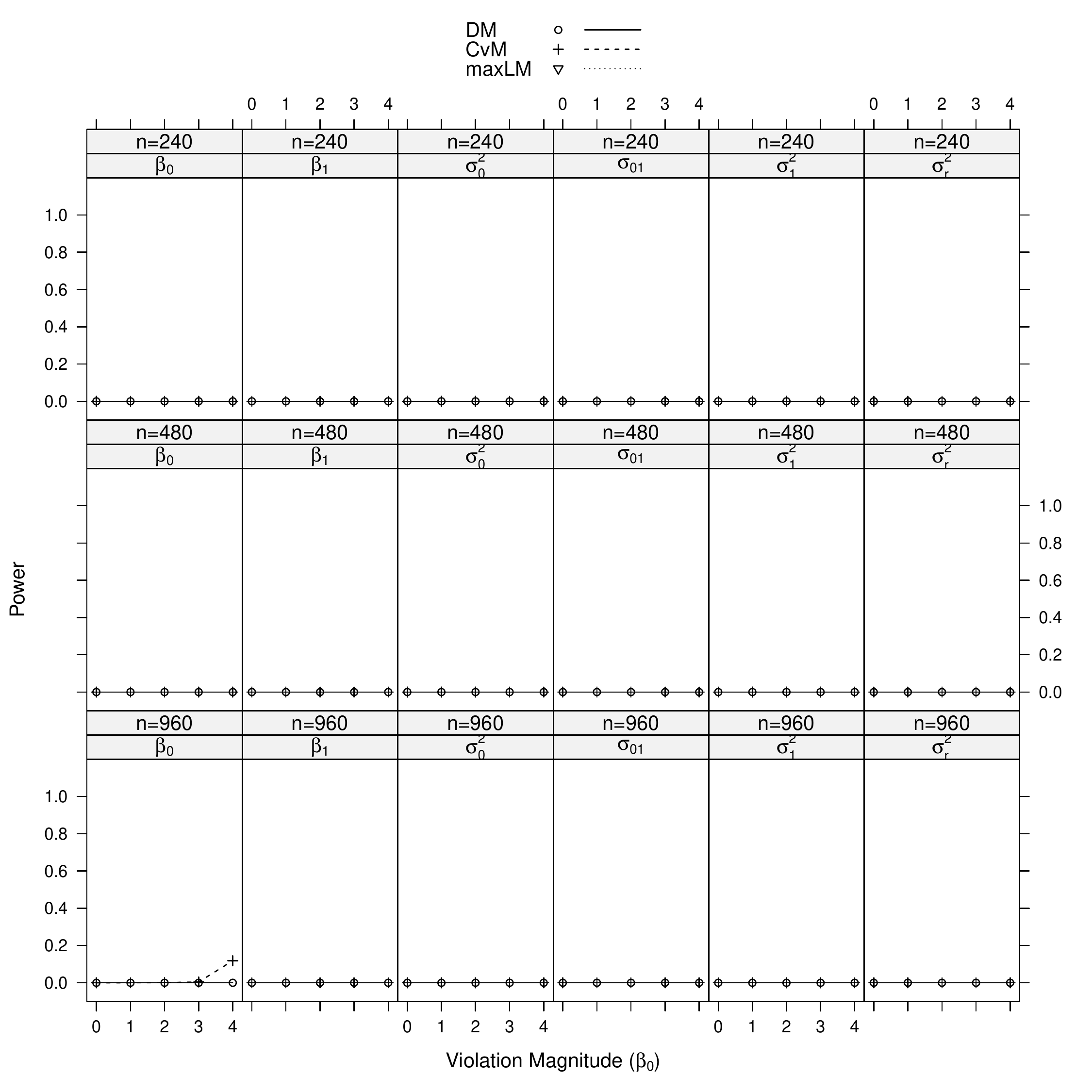}
\end{figure}

\begin{figure}[H]
\caption{\scriptsize{Simulated power curves for $\text{CvM}, \text{DM}, \text{maxLM}$ across parameter change of
         0--4 asymptotic standard error. 
         The changing parameter is $\beta_{1}$. Panel labels denote
         the parameter being tested along with sample size.}}
\label{fig:sim12res}
\includegraphics{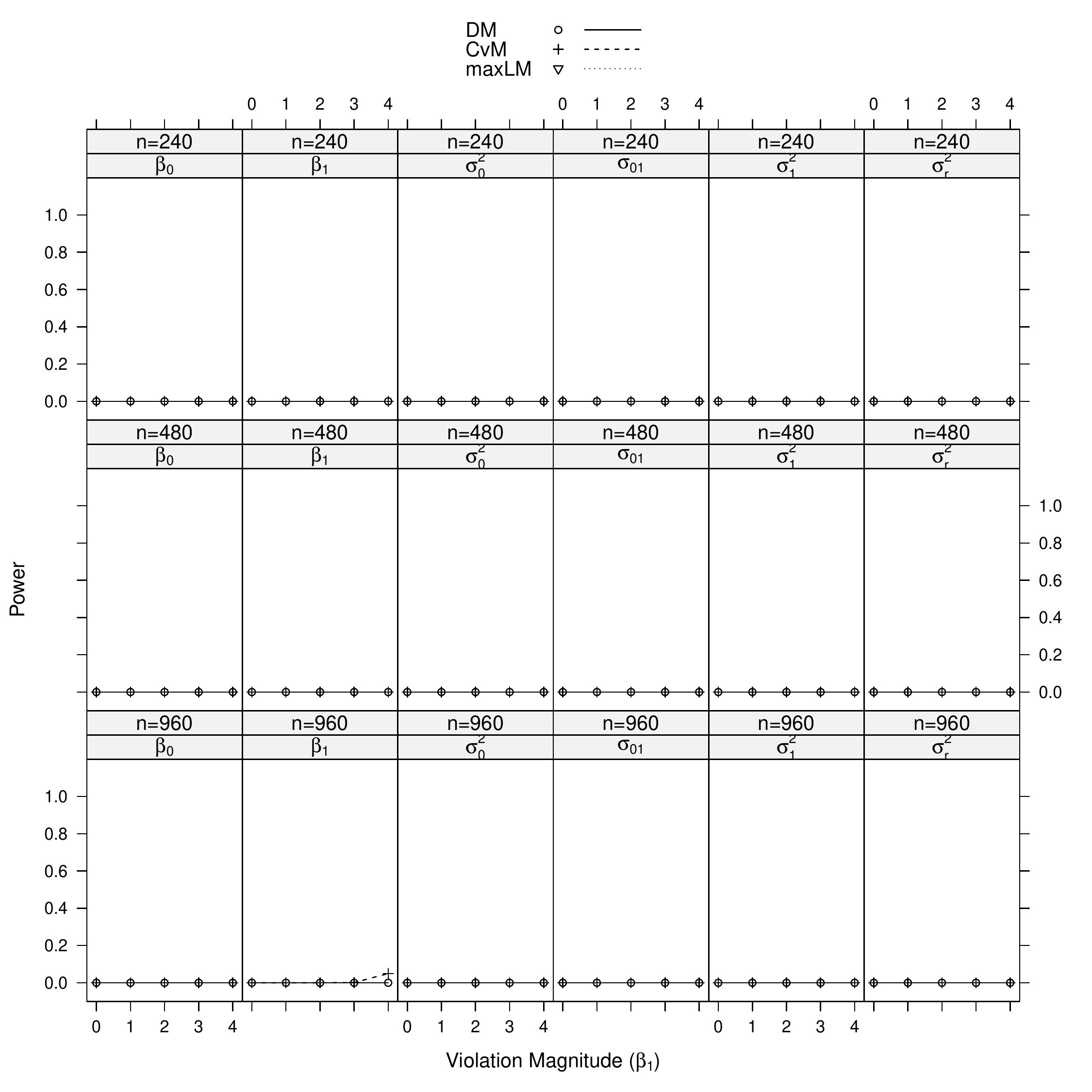}
\end{figure}

\begin{figure}[H]
\caption{\scriptsize{Simulated power curves for 
$\text{SN}$,  across parameter change of
         0--4 standard errors. 
         The changing parameter is $\beta_{0}$. Panel labels denote
         the parameter being tested along with sample size.}}
\label{fig:sim13res}
\includegraphics{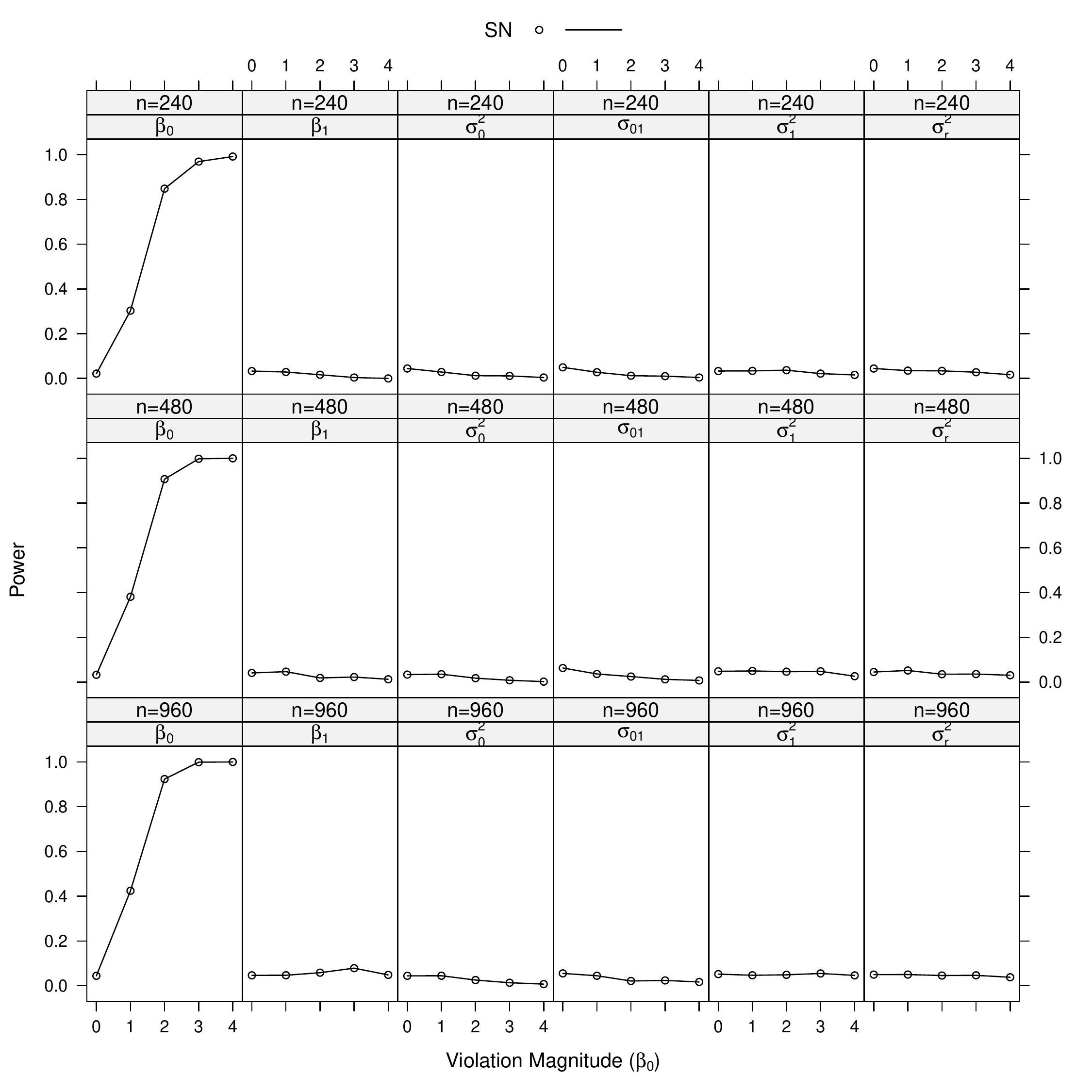}
\end{figure}

\begin{figure}[H]
\caption{\scriptsize{Simulated power curves for 
$\text{SN}$,  across parameter change of
         0--4 standard errors. 
         The changing parameter is $\beta_{1}$. Panel labels denote
         the parameter being tested along with sample size.}}
\label{fig:sim14res}
\includegraphics{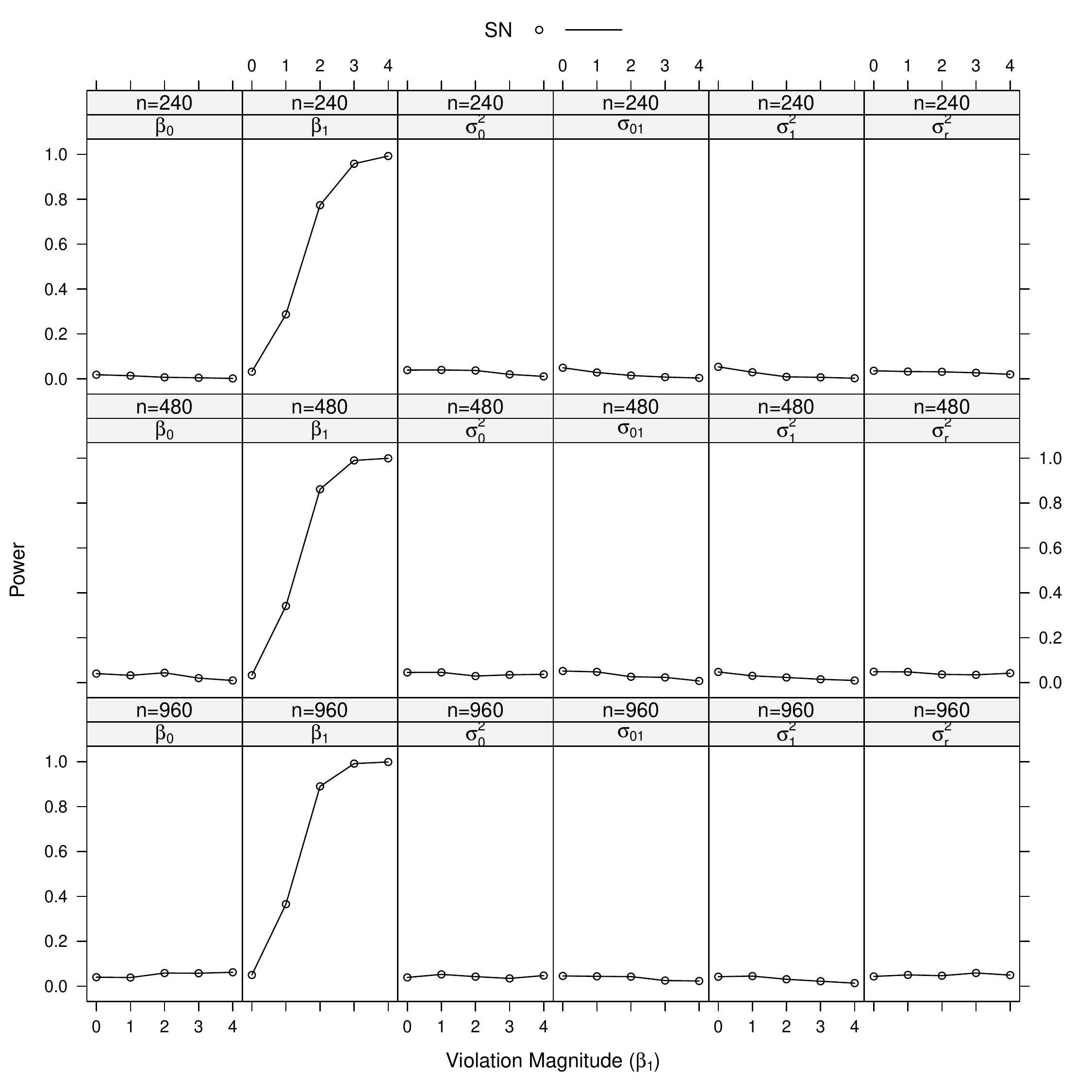}
\end{figure}

The full simulation results for $\beta_{0}$  and $\beta_{1}$ are demonstrated 
in Figures~\ref{fig:sim11res} to~\ref{fig:sim14res}. These figures are arranged similarly, with the first two showing results for the traditional statistics (which assume independence of scores) and the second two showing results for the self-normalized statistic. In all four figures, the panel titles indicate the sample size and the tested parameter, and the y-axis indicates power
(using $\alpha = 0.05$).  Figures~\ref{fig:sim11res} and~\ref{fig:sim13res} show results when $\beta_0$ exhibits instability, and Figures~\ref{fig:sim12res} and~\ref{fig:sim14res} show results when $\beta_1$ exhibits instability.

Looking at the first two figures, it is clear that the traditional statistics exhibit virtually no power in this scenario. The scores' lack of independence lead to this problem. 
On the contrary, the self-normalized results in Figures~\ref{fig:sim13res} and~\ref{fig:sim14res} exhibit good power for the parameter that is truly changing. The power is monotonic with $d$ and increases with sample size.  When $d \geq 3$, even the smallest sample size of $240$ demonstrates high power (power greater than 95\%). Additionally, we can see the power is slightly higher for parameter $\beta_0$ compared to $\beta_1$. This phenomenon was also observed in \citeA{wang182}.

Tables~\ref{tab:sim1} and~\ref{tab:sim2} show the exact numbers underlying Figures~\ref{fig:sim13res} and~\ref{fig:sim14res}, which is especially useful for examining Type I error rates (at $d=0$). Specifically, Type I error is equal to power when $d=0$ (when parameters do not change) or when tests are applied to parameters that do not change. We can see the Type I error rates are generally below or approaching 5\%. These results provide some evidence that self-normalized, score-based tests can be applied to models whose scores exhibit dependence.


\begin{table*}
 \caption{Simulated power for self-normalized score-based test across three sample sizes 
 $n$, five magnitudes of parameter change from 0 to 4 times of asymptotic standard error of 
  true changing parameter $\beta_0$, 
and six subsets of tested parameters. See Figure~\ref{fig:sim13res} for a visualization. }
  \label{tab:sim1}
  \begin{center}
  \begin{tabular}{lccrrrrr}
 \hline
Observation & Tested Parameter & Statistic & 0 & 1 & 2 & 3 & 4 \\ \hline
n=240 & $\beta_{0}$    & SN   &   2.2 &  30.3 &  84.8 &  96.9 &  99.2 \\
      & $\beta_{1}$    & SN   &   3.3 &   2.9 &   1.6 &   0.4 &   0.0 \\
      & $\sigma_{0}^2$ & SN   &   4.4 &   2.9 &   1.2 &   1.1 &   0.4 \\
      & $\sigma_{01}$  & SN   &   5.0 &   2.8 &   1.2 &   1.0 &   0.4 \\
      & $\sigma_{1}^2$ & SN   &   3.3 &   3.4 &   3.7 &   2.2 &   1.6 \\
      & $\sigma_{r}^2$ & SN   &   4.4 &   3.5 &   3.4 &   2.8 &   1.7 \\
n=480 & $\beta_{0}$    & SN   &   3.3 &  38.1 &  90.7 &  99.8 & 100.0 \\
      & $\beta_{1}$    & SN   &   4.1 &   4.7 &   1.9 &   2.3 &   1.3 \\
      & $\sigma_{0}^2$ & SN   &   3.4 &   3.5 &   1.8 &   0.8 &   0.2 \\
      & $\sigma_{01}$  & SN   &   6.3 &   3.6 &   2.5 &   1.2 &   0.7 \\
      & $\sigma_{1}^2$ & SN   &   4.8 &   5.0 &   4.7 &   4.8 &   2.6 \\
      & $\sigma_{r}^2$ & SN   &   4.5 &   5.2 &   3.5 &   3.6 &   3.1 \\
n=960 & $\beta_{0}$    & SN   &   4.4 &  42.4 &  92.4 &  99.9 & 100.0 \\
      & $\beta_{1}$    & SN   &   4.7 &   4.7 &   5.8 &   7.9 &   4.8 \\
      & $\sigma_{0}^2$ & SN   &   4.4 &   4.5 &   2.5 &   1.3 &   0.7 \\
      & $\sigma_{01}$  & SN   &   5.5 &   4.5 &   2.1 &   2.4 &   1.7 \\
      & $\sigma_{1}^2$ & SN   &   5.2 &   4.7 &   4.9 &   5.5 &   4.6 \\
      & $\sigma_{r}^2$ & SN   &   5.0 &   5.0 &   4.6 &   4.7 &   3.8 \\ \hline\end{tabular}
\end{center}
\end{table*}

\begin{table*}
 \caption{Simulated power for self-normalized score-based test across three sample sizes 
 $n$, five magnitudes of parameter change from 0 to 4 times of asymptotic standard error of 
 true changing parameter $\beta_1$, 
and six subsets of tested parameters. See Figure~\ref{fig:sim14res} for a visualization. }
  \label{tab:sim2}
  \begin{center}
  \begin{tabular}{lccrrrrr}
 \hline
Observation & Tested Parameter & Statistic & 0 & 1 & 2 & 3 & 4 \\ \hline
n=240 & $\beta_{0}$    & SN   &  1.9 &  1.4 &  0.7 &  0.5 &  0.2 \\
      & $\beta_{1}$    & SN   &  3.2 & 28.7 & 77.4 & 95.9 & 99.3 \\
      & $\sigma_{0}^2$ & SN   &  3.9 &  4.0 &  3.8 &  2.0 &  1.1 \\
      & $\sigma_{01}$  & SN   &  5.0 &  2.9 &  1.5 &  0.8 &  0.4 \\
      & $\sigma_{1}^2$ & SN   &  5.4 &  3.0 &  0.9 &  0.7 &  0.3 \\
      & $\sigma_{r}^2$ & SN   &  3.6 &  3.3 &  3.2 &  2.7 &  2.0 \\
n=480 & $\beta_{0}$    & SN   &  4.0 &  3.2 &  4.4 &  2.0 &  0.9 \\
      & $\beta_{1}$    & SN   &  3.3 & 34.1 & 86.1 & 99.0 & 99.9 \\
      & $\sigma_{0}^2$ & SN   &  4.5 &  4.6 &  2.9 &  3.4 &  3.7 \\
      & $\sigma_{01}$  & SN   &  5.1 &  4.8 &  2.6 &  2.3 &  0.7 \\
      & $\sigma_{1}^2$ & SN   &  4.7 &  3.0 &  2.3 &  1.5 &  0.9 \\
      & $\sigma_{r}^2$ & SN   &  4.8 &  4.8 &  3.6 &  3.4 &  4.2 \\
n=960 & $\beta_{0}$    & SN   &  4.0 &  3.9 &  5.9 &  5.8 &  6.2 \\
      & $\beta_{1}$    & SN   &  5.0 & 36.5 & 89.0 & 99.2 & 99.9 \\
      & $\sigma_{0}^2$ & SN   &  3.9 &  5.3 &  4.3 &  3.5 &  4.7 \\
      & $\sigma_{01}$  & SN   &  4.6 &  4.4 &  4.3 &  2.5 &  2.3 \\
      & $\sigma_{1}^2$ & SN   &  4.2 &  4.5 &  3.1 &  2.2 &  1.4 \\
      & $\sigma_{r}^2$ & SN   &  4.3 &  5.0 &  4.7 &  5.9 &  4.9 \\ \hline\end{tabular}
\end{center}
\end{table*}


\section{Application}
\citeA{wang21} provided an application of score-based tests to linear mixed models, studying heterogeneity in the relationship between socioeconomic status and math test scores. They showed that the tests could detect heterogeneity in both fixed effects and random effect variances, when the heterogeneity occurred with respect to a level-2 auxiliary variable. Those tests were restricted to level-2 auxiliary variables, though, due to the dependence issue described throughout this chapter. In this section, we use self-normalization to conduct similar tests with respect to a level-1 auxiliary variable.

\subsection{Method}
We use the {\em bdf} dataset \cite{sni11} from R package \pkg{mlmRev} \cite{mlmRev}. The dataset contains language test scores of 2,287 students from 131 schools. Variables in the dataset include students' verbal IQs, along with language test scores at two timepoints (``pre'' and ``post'').

The aim of the current analysis is to determine how students' language scores at time 2 
(denoted as \emph{langPOST} in the dataset) are associated with 
their language scores at time 1, along with their verbal IQs.  It
is plausible that the relationship between time 1 and 2 language scores
differ for students with different verbal IQs.
As illustrated in previous research \cite{wang21}, such an interaction can be insignificant due 
to heterogeneity in random effect variances.  Furthermore, because both covariates are at level 1, we have dependence in the scores. 
We use the self-normalized score test to study heterogeneity in mixed model parameters with respect to verbal IQ. The model specifications are shown via code in Figures~\ref{modelfit} and~\ref{modelest}.

\begin{figure}
  \caption{Code for including an interaction directly in the linear mixed model.}
  \label{modelfit}
\begin{Schunk}
\begin{Sinput}
> library("mlmRev")
> library("lmerTest")
> data("bdf")
> m1 <- lmer(langPOST ~ IQ.verb * langPRET + (1 | schoolNR), data = bdf,
+            REML = FALSE)
\end{Sinput}
\end{Schunk}
\end{figure}

\subsection{Results}
The most common approach to testing the interaction between verbal IQ and time 1 scores involves including the interaction directly in the mixed model, as shown in Figure~\ref{modelfit}.  
In estimating this model, we observe that the coefficient for the interaction term is not significant ($t(2264) = -1.182, p = 0.24$). But the significance test for the interaction
might be impacted by variance/covariance heterogeneity in random effects \cite{wang21}. Thus, 
we use the self-normalized score-based tests to distinguish
between the level-1 interaction and variance heterogeneity. 

To conduct the score-based tests, we first fit a model with time 1 scores as the only covariate,
as shown in Figure~\ref{modelest}. The score test is then carried out with verbal IQ as the auxiliary variable. Because verbal IQ is a level 1 covariate, 
the traditional score-based tests are not appropriate due to dependence of the scores.

\begin{figure}
  \caption{Code for fitting a model with only one fixed effect.}
  \label{modelest}
\begin{Schunk}
\begin{Sinput}
> m2 <- lmer(langPOST ~ langPRET + (1 | schoolNR), data = bdf,
+            REML = FALSE)
\end{Sinput}
\end{Schunk}
\end{figure}

\begin{figure}
  \caption{Empirical statistics process based on self-normalization approach.}
  \label{fig:fluctuate}
\includegraphics{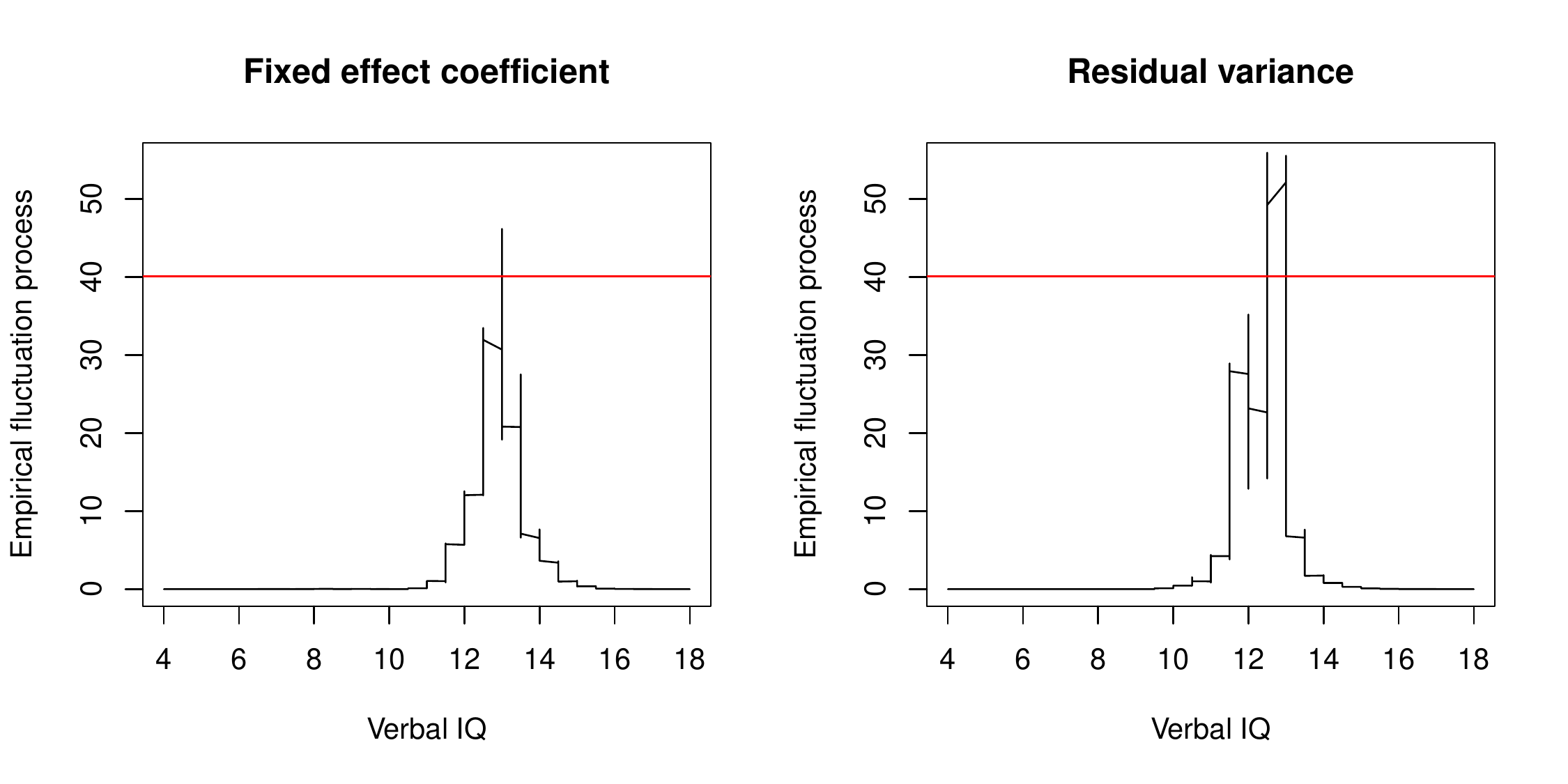}
\end{figure}

We compute a self-normalized score test statistic separately for the fixed effect of time 1 test scores and for the residual variance, 
 and the corresponding code is shown in Figure~\ref{fig:fluctuate1}. 
The empirical statistics' fluctuation 
processes are demonstrated in Figure~\ref{fig:fluctuate}, 
where the left panel is for the fixed effect and the right panel is for the residual variance.  The black line shows the value of the $SN$ statistic at different values of verbal IQ, with the maximum (the highest point) being the official value of the test statistic. The red line indicates 
the critical value, where is obtained from Table 1 \cite{shao10} based on the number of testing parameters and alpha level.  Because the black line crosses the red line in both cases, we conclude that both the fixed effect of time 1 scores and the residual variance fluctuate with verbal IQ.

Because both black lines of Figure~\ref{fig:fluctuate} peak around verbal IQs of 13, we can conclude that students with verbal IQs below 13 have different parameter values than students with verbal IQs above 13.  This leads us to estimate two separate models, one for students with IQs less than or equal to 13, and one for students with IQs greater than 13. In estimating these separate models, we find that the fixed effect of time 1 scores is estimated at $0.89$ and $0.69$ for students with verbal IQs lower than 13 and higher than 13, respectively. The residual variance for students with verbal IQs below 13 is estimated at 33, while the residual variance for students with verbal IQs above 13 is estimated at 17.

These numerical results indicate that, while the association between time 1 and time 2 scores is somewhat lower for students with high verbal IQ, those students have lower residual variance and generally obtain high test scores. On the other hand, while time 1 scores are more helpful for predicting time 2 scores of students with low verbal IQ, those students also exhibit more residual variability in their scores. These types of results might help practitioners adopt different strategies for different types of students.

\begin{figure}
 \caption{Code to generate self-normalized statistics}
 \label{fig:fluctuate1}
\begin{Schunk}
\begin{Sinput}
>   library("strucchange")  
>   library("merDeriv")  
>   source("estfun.lmerMod.R")
>   source("mzall.R")
>   psi1 <- estfun.lmerMod(m2)
>   mz_gefp1  <- as.matrix(psi1[,2])
>   mz_gefp2  <- as.matrix(psi1[,4])
>   statsval1 <- calc.self.stats.cont(mz_gefp1, bdf$IQ.verb)
>   statsval2 <-  calc.self.stats.cont(mz_gefp2, bdf$IQ.verb)
\end{Sinput}
\end{Schunk}
\end{figure}

\section{General Discussion}
In this chapter, we demonstrated that the self-normalization approach can be used to carry out score-based tests in situations where scores exhibit dependence. This allows researchers to apply score-based tests to mixed models and other models where observations are not independent. It extends related work on score-based tests for mixed models \cite{fok15,wang21,wangra22}, allowing us to test for heterogeneity with respect to level-1 auxiliary variables.
While this chapter provides evidence that self-normalization is promising for score-based tests, it leaves open many issues for future work. We briefly describe some issues in the sections below.

\subsection{Extensions of Self-Normalization}
In this chapter, we used self-normalization to test one model parameter at a time. We also used the statistic that was previously proposed by \citeA{zhang11}. In future work, it would be worthwhile to test multiple parameters at a time and to use self-normalization in other statistics.

If we test for multiple parameters at at time, it is unclear whether we should first decorrelate the cumulative scores via $\hat{\bm I}$, prior to self-normalization. This is because the self-normalization procedure can potentially address both row-wise (cases) and column-wise (parameters) score dependence. However, because the information matrix is often obtained analytically as opposed to numerically, we may see improved power from a self-normalization test that first decorrelates via $\hat{\bm I}$.

Additionally, while we focused on the $SN$ statistic from Equation~\eqref{eq:dm2}, it appears that $\bm{V}_n(k)$ could also replace 
$\hat{\bm I}$ in the CvM statistic of Equation~\eqref{eq:transfer4}. The $\bm{V}_n(k)$ matrix may similarly replace $\hat{\bm I}$ in other score-based statistics, leading to a family of self-normalized, score-based tests that is similar to the traditional family of score-based tests. The family of self-normalized tests could be applied to a wider variety of statistical models.

\subsection{Weighted Statistics}
\citeA{shao10, shao15} discuss how we can include weights in the $SN$ statistic, which would be useful for carrying out score-based tests with respect to ordinal variables \cite{merfanzei}. Specifically, let $w(t), t \in [0,1]$ be the weight function, then the general form of the weighted $SN$ statistic is 
\begin{equation}
SN_w = \sup_{1 \leq k \leq n-1} w(k/n)\ \bm{T}_n(k)^{T} \bm V_{n}^{-1}(k) 
             \bm {T}_n(k).
\end{equation}
\citeA{merfanzei} described weights that allow for score-based tests with respect to ordinal auxiliary variables, which could be inserted into the above equation as 
\begin{equation}
SN_{ord} = \sup_{k \in k_1, \ldots, k_{m-1}}\left \{\frac{k}{n}\left (1-
\frac{k}{n}\right )\right\}^{-1}
 \bm{T}_n(k)^{T} \bm V_{n}^{-1}(k) \bm {T}_n(k), 
\end{equation}
where $m$ represents the number of ordinal levels in the auxiliary variable. This means that the self-normalized statistics could flexibly handle many aspects of traditional score-based statistics.

\subsection{Computation}
Assuming that the ideas in the previous two subsections work as expected, we imagine an implementation similar to the functionality of the \pkg{strucchange} package \cite{strucchange}. This would provide users with score-based functionality for a wider variety of mixed modeling situations. With this in mind, our code is currently far from optimal and could be improved in a variety of manners. Here, we consider recursive computations of the self-normalized test statistic.

Recursive computations of the $\bm{V}_n(k)$ matrix from Equation~\eqref{eq:v} may be helpful, because that matrix must be computed and inverted for each value of $k = 1,\ldots,(n-1)$. The inversion is not a major problem if we test one parameter at a time (as we did in this chapter), but it could become computationally slow if we test many parameters at a time. 
To speed this up, it appears possible to write $\bm{V}_n(k)$ as a function of $\bm{V}_n(k-1)$ using identities from \citeA<>[also see the helpful Wikipedia entry on Welford's online algorithm]{wel62}. That result might then be plugged in to the Sherman-Morrison formula, allowing us to compute $\bm{V}_n^{-1}(k)$ from $\bm{V}_n^{-1}(k-1)$. This could enable us to invert the $q \times q$ matrix for a single value of $k$, then obtain subsequent inverses through simple matrix multiplications. Of course, the full derivations remain to be worked out.

\subsection{Summary}
Given the flexibility of the self-normalization approach and the existing applications of score-based tests, the combination of these two methods appears fruitful. We expect that further development of the ideas presented here can provide enhanced score-based tools to researchers in the social sciences and beyond.

\section*{Computational Details}
All results were obtained using the \proglang{R}~system for 
statistical computing \cite{r22},
version~4.0.3, especially including packages \pkg{lme4}~1.1-26 \cite{lme4} for model estimation, \pkg{merDeriv}~0.2-4 \cite{wang182} for score computations, and \pkg{strucchange}~1.5-2 \cite{strucchange} for score-based tests. Code for reproducing our results is available at \url{http://semtools.r-forge.r-project.org/}.

\proglang{R}~and the aforementioned packages are
freely available under the General Public License~2 from the
Comprehensive \proglang{R} Archive Network 
at \url{https://CRAN.R-project.org/}.

\bibliography{refs}

\end{document}